\documentclass[sigconf,edbt]{acmart-edbt2025}

\def\BibTeX{{\rm B\kern-.05em{\sc i\kern-.025em b}\kern-.08em
    T\kern-.1667em\lower.7ex\hbox{E}\kern-.125emX}}

\usepackage{booktabs} 

\setcopyright{rightsretained}

\acmDOI{}

\acmISBN{978-3-98318-097-4}

\usepackage[utf8]{inputenc}
\usepackage[toc,page]{appendix}
\usepackage[export]{adjustbox}
\usepackage{algorithm} 
\usepackage[noend]{algpseudocode}
\usepackage{amsmath}
\usepackage{tablefootnote}
\newenvironment{defn}[1][Definition]{%
  \textbf{#1} %
}{%
  \par
}
\usepackage{tikz}
\usepackage{xcolor}

\usepackage{todonotes}

\def\thickhline{\noalign{\hrule height 1.25pt}}

\title{TransClean: Finding False Positives in Multi-Source Entity Matching under Real-World Conditions via Transitive Consistency}

\author{Fernando De Meer Pardo}
\email{fernando.demeerpardo@uzh.ch}
\affiliation{
    \institution{University of Zurich}
    \city{Zurich}
    \country{Switzerland}
}
\affiliation{
    \institution{Zurich University of Applied Sciences}
    \city{Winterthur}
    \country{Switzerland}
}

\author{Branka Hadji Misheva}
\email{branka.hadjimisheva@bfh.ch}
\affiliation{
    \institution{Bern University of Applied Sciences}
    \city{Bern}
    \country{Switzerland}
}

\author{Martin Braschler}
\email{martin.braschler@zhaw.ch}
\affiliation{
    \institution{Zurich University of Applied Sciences}
    \city{Winterthur}
    \country{Switzerland}
}

\author{Kurt Stockinger}
\email{kurt.stockinger@zhaw.ch}
\affiliation{
    \institution{Zurich University of Applied Sciences}
    \city{Winterthur}
    \country{Switzerland}
}
\renewcommand{\shortauthors}{De Meer Pardo et al.}

\begin{document}

\begin{abstract}

We present TransClean, a method for detecting false positive predictions of entity matching algorithms under real-world conditions characterized by large-scale, noisy, and unlabeled multi-source datasets that undergo distributional shifts. TransClean is explicitly designed to operate with multiple data sources in an efficient, robust and fast manner while accounting for edge cases and requiring limited manual labeling. TransClean leverages the \textit{Transitive Consistency} of a matching, a measure of the consistency of a pairwise matching model $f_\theta$ on the matching it produces $G_{f_\theta}$, based both on its predictions on \textit{directly evaluated record pairs} and its predictions on \textit{implied record pairs}. TransClean iteratively modifies a matching through gradually removing false positive matches while removing as few true positive matches as possible. In each of these steps, the estimation of the \textit{Transitive Consistency} is exclusively done through model evaluations and produces quantities that can be used as proxies of the amounts of true and false positives in the matching while not requiring any manual labeling, producing an estimate of the quality of the matching and indicating which record groups are likely to contain false positives. In our experiments, we compare combining TransClean with a naively trained pairwise matching model (DistilBERT) and with a state-of-the-art end-to-end matching method (CLER) and illustrate the flexibility of TransClean in being able to detect most of the false positives of either setup across a variety of datasets. Our experiments show that TransClean induces an average +24.42 F1 score improvement for entity matching in a multi-source setting when compared to traditional pair-wise matching algorithms.

\end{abstract}

\maketitle
\section{Introduction}\label{Section: Introduction}
\emph{Entity Matching} (EM) describes the process in which data records originating from different sources are compared to find those that refer to the same \emph{real-world entities}. EM is a vital part of any \textit{Extract-Transform-Load (ETL)/data integration} process, as it enables the joint use of all available data items relating to each real-world entity when such items originate from different sources. Successful EM systems allow for the merging of different data sources into unified repositories and prevent inaccuracies caused by duplicate records. EM, itself ironically referred to via multiple names such as \emph{Record Linkage} (RL), \emph{Entity Resolution} (ER) and \emph{Data Integration/Deduplication}, has been studied since the 1940s \cite{Dunn1946-xz, Fellegi1969ATF} and has been classically tackled via rule-based algorithms and heuristics \cite{Hassanzadeh2009FrameworkFE, Magellan2016, JedAI2020, OverviewofEnd2EndERBigData2020}. In recent years, however, most state-of-the-art EM methods rely on machine learning (ML), most notably Transformer models \cite{DeepEMChallengesOpportunities2021, Li2023, Brunner2020, CLER2024}.

ML-based methods frame EM as a \emph{binary classification} problem by classifying pairs of records as either \texttt{Match} or \texttt{NoMatch} and are usually compared to each other based exclusively on their performance on this task. However, based on individual pairwise predictions, entity matching models also implicitly define groupings of records through transitive (implicit) links. These transitive relationships play a crucial role in multi-source settings and we will discuss them in detail in Section \ref{Section: Methodology}.

\begin{figure}[h!t]
  \centering
  \includegraphics[width=\linewidth, keepaspectratio]{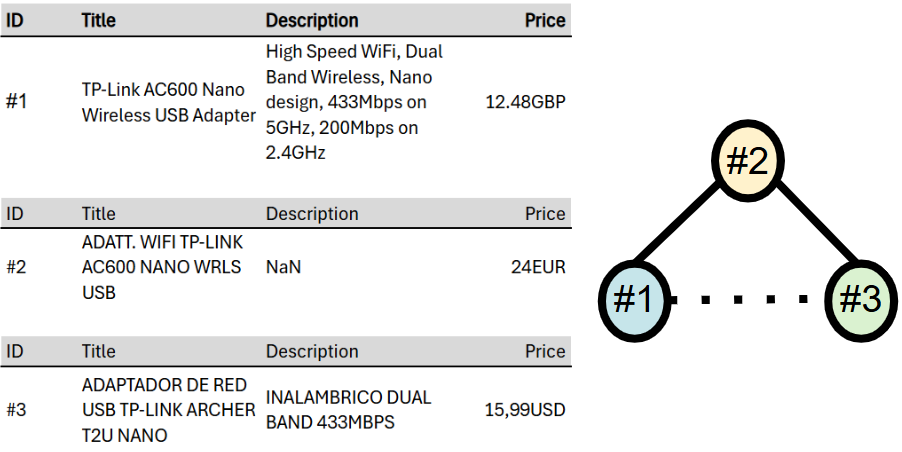}
  \caption{Example of a matching involving three records from three different data sources. The matches ($r_{\#1}$, $r_{\#2}$) and ($r_{\#2}$, $r_{\#3}$) imply also the transitive match ($r_{\#1}$, $r_{\#3}$) even if $r_{\#1}$ and $r_{\#3}$ have not been directly paired.}
  \label{fig:TransMatchExample}
\end{figure}

Consider, for instance, the matching of records $r_{\#1}$, $r_{\#2}$ and $r_{\#3}$ from three different data sources as shown in Figure \ref{fig:TransMatchExample}. If an EM method predicts pairs ($r_{\#1}$,$r_{\#2}$) and ($r_{\#2}$,$r_{\#3}$) as matches, then it is implying the match ($r_{\#1}$,$r_{\#3}$) as well (see dotting line in Figure \ref{fig:TransMatchExample}). Such matches have been referred to as \emph{transitive} matches \cite{DBLP:conf/edbt/PardoLGNNMBS25} and can be present in any matching setting - even in the standard setting of matching records from only two different sources. Transitive matches are, however, most problematic in the multi-source matching case, due to the greater numbers of candidate pairs being considered and the propagating effect of false positive predictions in terms of wrongly linking different record groups, which leads to large numbers of false positive transitive matches \cite{DBLP:conf/edbt/PardoLGNNMBS25}.

In the real-world, EM tasks involve numerous large unlabeled datasets which make evaluating the performance of different EM methods and comparisons among them unfeasible in practice, as verifying the faithfulness of a given set of pairwise matches requires manually labeling them one by one, which is extremely time consuming when dealing with millions of record pairs. Effectively, the evaluation of a given matching across unlabeled data sources is in itself as challenging as producing the matching itself, if not even more so. Comparisons between different matchings can be attempted via analyzing, for example, statistics of the produced record groups such as the size distribution, the number of pairwise predictions per group, etc. Such comparisons, however, are not conclusive and cannot alone pinpoint the strengths and weaknesses of each EM method unless extensive manual labeling is also involved.

This difficulty has created a significant gap between EM research and real-world projects. On one hand, EM research is focused on improving state-of-the-art black box pairwise prediction models where ground truth matches are available. On the other hand, in real-world projects, EM tasks are usually tackled through heuristics on a case-by-case basis \cite{TailoringEM2020, DataIntCleanDedupl2022}, where ground truth matches are typically not available or very expensive to get. The narrowing of this gap can only be achieved through developing EM methods that account for several criteria such as:
\begin{itemize}
    \item \textbf{Manual labeling:} Only a reduced number of candidate record pairs can be expected to be manually labeled by human annotators. As a task, manual labeling is inherently labor-intensive and monotonous, often leading to annotator fatigue, which negatively impacts labeling accuracy over time \cite{PANDEY2022102772}.
    
    \item \textbf{Run time:} EM methods need to produce matchings in a reasonable time, the complexity of every step of an EM pipeline needs to be carefully accounted for.
    
    \item \textbf{Correctness assurance:} Black box EM methods should provide some form of additional evidence on the correctness of the matching produced beyond the predictions themselves, so that extensive manual labeling can be avoided.

\end{itemize}

Our aim is to add to this line of research by presenting the concept of \emph{Transitive Consistency}, a measure of the consistency of a pairwise matching model $f_\theta$ based on its predictions on the \textit{transitive matches} of its produced matching $G_{f_\theta}$. We also introduce \textit{TransClean}, a technique that leverages the Transitive Consistency to identify and remove false positive predictions through a series of iterative steps and accounts for all the criteria listed previously. Our \underline{main contributions} are as follows:

\begin{enumerate}
    \item We introduce the concept of Transitive Consistency and TransClean, a method that can be combined with any pairwise matching model and aims to detect the false positive predictions of a given matching. TransClean is inspired by a large real-world use case where entities of multiple sources need to be matched. Hence, TransClean is explicitly designed for the matching of multiple, large and unlabeled datasets in an efficient, robust and fast manner while accounting for edge cases and requiring little manual labeling.

    \item In our experiments, we combine TransClean both with a naively trained pairwise matching model (DistilBERT) and a state-of-the-art EM method (CLER \cite{CLER2024}) across a variety of datasets and illustrate TransClean's ability to detect large proportions of the false positives produced by both models.

    \item We illustrate the advantages of using TransClean over using exclusively traditional pairwise matching algorithms. Specifically, we show how the Transitive Consistency, calculated via the predictions on transitive matches, can be used as an observable proxy of the quality of a matching and how its evolution during the different steps of TransClean indicates a gradual improvement of the matching, eliminating the need to perform extensive manual labeling. Our experiments show that TransClean induces an average +24.42 F1 score improvement for entity matching in a multi-source setting when compared to traditional pair-wise matching algorithms.

\end{enumerate}
The manuscript is organized as follows. Section \ref{Section: Related Work} discusses the related work on EM with ML methods. Section \ref{Section: Real-world EM conditions} presents a discussion on the challenges that arise in real-world EM tasks and the issues associated with using exclusively pairwise matching models. Section \ref{Section: Methodology} presents the definition of Transitive Consistency and the different steps that make up TransClean. Section \ref{Section:Experiments} discusses the setup and results of the different entity matching experiments we run across multiple datasets. Finally, Section \ref{Section: Conclusions and Future Work} presents our conclusions along with an outlook on future work. 

\section{Related Work}\label{Section: Related Work}

Current state-of-the-art EM methods employ ML models based on variations of the Transformer architecture \cite{2017AttentionTransformer}. Classical EM methods based on combinations of hand-crafted heuristics \cite{Magellan2016, JedAI2020} have been overtaken by ML-based ones due to the ability of the former at overcoming common EM challenges such as noisy and heterogeneous records (e.g. with missing data, misplaced attributes etc.) by treating them as instances of textual data. Through fine-tuning, pre-trained Transformer models are able to learn complex mappings from sets of manually labeled pairs.

Following this line of thought, multiple works present different training optimizations, such as pseudo-labeling, data augmentation and contrastive learning, that aim to improve the performance of pairwise pre-trained Transformer classifiers \cite{Brunner2020, Li2023, ALMSER2021,  miao2021rotom, Peeters2022SupervisedCL, wang2022sudowoodo, Yao2022EntityRW, CollaborEM2023}, often paired with Active Learning techniques \cite{DistributedRepTuplesER2018, Fu2019EndtoEndMM, LRDeepERTrAL2019, brunner2019entity, Li2020, HierMatchNetER2020, TailoringEM2020,AutoEM2019, ComprehensiveBenchFrALEM2020, BattleshipApproachLowResEM2023} and Large Language Models (LLMs) \cite{2022FoundationModels, 2023EMwLLMs, 2025DeepDiveCrossEM, 2024MatchCompareSelect}\footnote{LLM-based works however present concerns regarding test data leakage during pre-training (as EM benchmark datasets are public and thus likely part of pre-training corpora) which question reported scores and generalization capabilities}. More recently, even quantum neural networks have been used for entity matching \cite{bischof2025hybrid}.

These methods are usually compared based on their performance on multiple benchmarks consisting of 2 data sources such as those published by the Database Group of the University of Leipzig \cite{LeipzigDatasetsEM2010} and the Magellan datasets \cite{magellandata} which have fixed train/val/test splits of record pairs. This evaluation methodology contributes to the research/real-world gap since in the former, no fixed sets of candidate pairs are available for matching, instead one has to select which subset of pairs to be evaluated by the pairwise matching model, a step commonly referred to as \emph{blocking}. 

Blocking is a key part of EM as the set of all possible pairs grows quadratically w.r.t. the number of records being matched, which makes the evaluation of all possible pairs unfeasible for large datasets. This is, however, ignored when using fixed splits of record pairs to evaluate the performance of different EM methods. There are works that focus exclusively on blocking, also framing it as a ML problem \cite{Autoblock2019, DLforBlockinginEM2021}.  Other works such as Sudowoodo \cite{wang2022sudowoodo} and CLER \cite{CLER2024} perform both blocking and matching while also considering labeling budgets, which make them more applicable to real-world scenarios.

All of the previous methods consider only the pairwise model predictions in their evaluations but, as we previously discussed, matching involves producing record groups which contain both the explicit and the implicit/transitive matches produced by a model. Previous works which also consider records groups and carry out matching in an agglomerative fashion include techniques such as \textit{Clustering} \cite{Hassanzadeh2009FrameworkFE, HolisticEC2016, FAMER2018, Saeedi2017, SaeediKG2018} and \textit{Progressive ER/EM} \cite{ProgressiveER2018, PayasYouGoER2013, ProgressiveER, ERonDemand2022}. Other works focus on model mistakes and propose methods to rank the most important model predictions for manual labeling \cite{2019TowardsIntER, RiskAwareHumMacCoopER2020, 2018ImprovingMachineER}. Such rankings, however, are of limited practical use in the matching of multiple large unlabeled datasets, since the amount of selected pairs to label may still be overwhelming to inspect. In addition, earlier approaches such as collective matching using relational constraints (\cite{bhattacharya2007collective, DomIndDataCLERelGraph2006, CollectiveERFamNet2017}), aimed to enforce consistency across record groups using joint inference. While less scalable, these methods inspired consistency-based ideas such as our Transitive Consistency framework.

We add to these previous lines of work by presenting TransClean, a technique that explicitly aims to discover false positive predictions present in a matching with a limited labeling effort and provides information of the quality of a matching without any human intervention.

\section{Real-world Matching Conditions}\label{Section: Real-world EM conditions}

In most real-world scenarios, a standard way to attempt to match several data sources with a pairwise matching model would be as follows:
\begin{enumerate}
    \item Initially, a portion of matches would be selected to be manually labeled.
    \item One may divide this set of labeled pairs into \texttt{train/val/test} splits, fine-tune with the \texttt{train} split and select a set of model weights based on the performance on the \texttt{val} split.
    \item Finally, evaluate the pairs of the \texttt{test} split to estimate the performance of the model on unseen pairs.
\end{enumerate}
However, when using the selected model to match the entire dataset (via evaluating the candidate pairs resulting from applying a blocking to all unlabeled records), the size of the matching produced can be expected to be of \textit{magnitudes higher} than that of the labeled subset of pairs previously used for fine-tuning. This effectively makes evaluating the matching unfeasible, as doing so requires manually checking all of its matches. This is especially problematic in real-world scenarios where significant distributional shifts can lead to large numbers of erroneously predicted matches, either due to the difficulty inherent to the records themselves, or to data artifacts that lead to edge cases.

Some characteristics of the matching produced, such as the number of records per group, can be used to detect which record groups are likely to have false positive predictions, which tend to be the biggest produced groups due to the agglutinative effect of false positive matches \cite{DBLP:conf/edbt/PardoLGNNMBS25}. This would, however, be quite a rudimentary methodology to detect false positives, since it requires manually inspecting all the matches of record groups in descending order, saving little in terms of manual labeling effort, and provides no information about the quality of smaller record groups which are likely the most numerous. \textbf{The challenge is then to develop a technique capable of effectively identifying a majority of the \textit{false positive predictions} of a given matching, which preserves the largest amount of \textit{true positive predictions}. Moreover, the technique should operate with a \textit{limited manual labeling budget} and provide some sort of assurance on the correctness of the unchecked record groups.}

\section{Methodology}\label{Section: Methodology}
Entity Matching requires considering not only the \emph{explicitly} predicted matches produced by a matching method, but also the \emph{implied} matches i.e. all the non-adjacent pairs of records belonging to the same \emph{component}\footnote{A component of an undirected graph is a connected subgraph that is not part of any larger connected subgraph. Here, and throughout the whole manuscript, we use the term \emph{components} to refer to the produced record groups of the matching given by $G_{f_\theta}$.} of the graph $G_{f_\theta} = (V, E_{f_\theta})$ whose nodes and edges represent records and predicted matches respectively\footnote{Throughout the manuscript we will refer to records as nodes and edges as matches, pairs or predictions interchangeably.}. This is because all records belonging to the same \textit{component} of $G_{f_\theta}$ are implied to represent the same real-world entity. The actual output of a matching is a series of \textit{cliques}\footnote{A clique is a subset of vertices of an undirected graph such that every two distinct vertices in the clique are adjacent. Throughout the manuscript we refer to cliques as record groups and vice versa.} even if not all pairs of vertices of $G_{f_\theta}$ belonging to the same component are adjacent. That is to say, a set of pairwise predictions implicitly leads to a \textit{label propagation} step in which all transitively connected records are assigned to the same group/entity. 

Our focus is the detection of \textit{false positives} of pairwise classifiers applied to the problem of multi-source Entity Matching. We aim to do this by analyzing the \textit{consistency} of each classifier/matching model $f_\theta$ with regards to its implicit/transitive matches which emanate from its own explicitly predicted matches (i.e. from $G_{f_\theta}$). 

In this section we illustrate how this consistency property can be leveraged by a graph cleanup method that efficiently detects false positives. Additionally, we show how the calculations involved in estimating this property produce quantities that indicate the quality of a given matching without having access to the ground truth.

\subsection{Transitive Consistency}

\begin{defn}[Definition. Transitive Match:]
    Let $G_{f_\theta} = (V,E_{f_\theta})$ be a graph where $V$ is the set of records to be matched and $E_{f_\theta}$ are the predicted matches $(r_i, r_j)$ of a pairwise matching model $f_\theta$. Two nodes $r_{i*}$ and $r_{j*}$ are said to form an \emph{implicit} or \emph{transitive match} $(r_{i*}, r_{j*})$ of graph $G_{f_\theta}$ if they are non-adjacent while belonging to the same component of $G_{f_\theta}$.
\end{defn}

Let us reconsider the example graph shown in Figure \ref{fig:TransMatchExample}. In this case, the nodes $r_{\#1}$ and $r_{\#3}$ are a transitive match since they are non-adjacent, i.e. there is no direct edge between them.

\vspace{.1cm}
\begin{defn}[Definition. Transitive Consistency:]
    Let $G_{f_\theta} = (V,E_{f_\theta})$ be a graph where $V$ is the set of records to be matched and $E$ is the set of predicted matches $(r_i, r_j)$ by a pairwise matching model $f_\theta$. We say that $f_\theta$ is  \emph{transitively consistent} in $G_{f_\theta}$ if for every transitive match $(r_{i*}, r_{j*})$ of $G_{f_\theta}$: $$f_\theta((r_{i*}, r_{j*})) = \texttt{Match}$$
\end{defn}    

In the example of records $r_{\#1}$, $r_{\#2}$ and $r_{\#3}$ shown in Figure \ref{fig:TransMatchExample}, a given pairwise matching model $f_\theta$ is transitively consistent if given that $f_\theta((r_{\#1}, r_{\#2})) = \texttt{Match}$ and $f_\theta((r_{\#2}, r_{\#3})) = \texttt{Match}$ then $f_\theta((r_{\#1}, r_{\#3})) = \texttt{Match}$ as well. Intuitively, the Transitive Consistency is a check of whether a pairwise matching model $f_\theta$ fully agrees on all the possible pairs in each component it has produced and if it does not, it points out in which record group/component the discrepancy arises.  

\begin{figure}[h!t]
  \centering
  \includegraphics[width=0.7\linewidth, keepaspectratio]{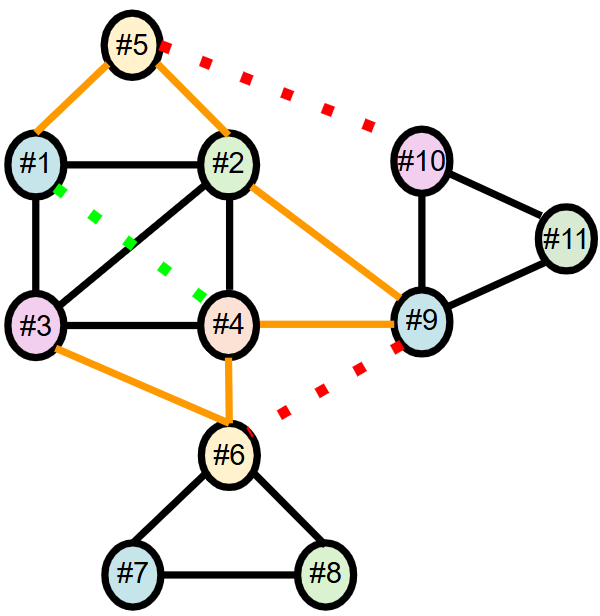}
  \caption{Example record group implied by a set of pairwise predictions between records $\{r_i\}_{i=1}^{11}$ originating from 5 different data sources. Black edges illustrate true positive predictions, orange edges false positive predictions, red dotted edges denote two of the \emph{transitive matches} (not all possible transitive matches are drawn) between wrongly linked records (due to the false positive predictions) and the green dotted edge denotes a \emph{transitive match} between correctly linked records.}
  \label{fig:Example_component_fig}
\end{figure}

Figure \ref{fig:Example_component_fig} illustrates an example record group implied by a set of pairwise predictions with some of its transitive matches. The records are from five different sources (indicated by different colors of the nodes). The transitive match ($r_{\#1}$, $r_{\#4}$) is a true match because it involves records transitively connected by a path made up exclusively of true positive predictions (shown as black edges). In turn,  ($r_{\#5}$, $r_{\#10}$) and ($r_{\#6}$, $r_{\#9}$) are non-matches because all the paths between said records include false positive predictions (shown as orange edges). In order for $f_\theta$ to be  \emph{transitively consistent} it would have to predict all possible pairs between records of the component as a $\texttt{Match}$.

Note that not all transitive matches $(r_{i*}, r_{j*})$ of each component need to be evaluated in order to check for the Transitive Consistency of a model since a single $\texttt{NoMatch}$ prediction suffices. Nonetheless, evaluating all transitive matches provides an estimation on the number of inconsistencies per record group and overall in the matching. We will refer to the transitive matches predicted as $\texttt{Match}$ as \emph{positive transitive predictions} and those
predicted as $\texttt{NoMatch}$ as \emph{negative transitive predictions}.

Contrary to the true and false positives of a matching $G_{f_\theta} = (V,E_{f_\theta})$, the calculation of the Transitive Consistency of a model $f_\theta$ depends only on model predictions and does not require any manual labeling since, by definition, it is independent of the ground truth. Negative transitive predictions are very likely in components/record groups with false positive matches, both because false positive predictions link unrelated records and because they tend to create larger record groups than true positives and thus produce more transitive matches to be evaluated. Because of this, positive and negative transitive predictions can be used as a proxy of true and false positives of a matching, as we will show in the experiments of Section \ref{Section:Experiments}. 

This property becomes more informative as more data sources are introduced, since each additional source contributes candidate pairs that, when predicted as matches, generate new transitive connections. The more transitive matches in a given record group, the more informative the Transitive Consistency of said group becomes, as it requires the agreement of the pairwise matching model on a bigger number of record pairs. This turns the matching of additional data sources, which has classically been a challenge due to the increased likelihood of erroneous pairwise predictions, into an advantage.

\subsection{TransClean}
In this section we present TransClean, a cleanup technique that leverages the Transitive Consistency of a pairwise matching model to identify false positives. 
\subsubsection{Introductory Description} \hfill\\
TransClean consists of three different steps which can be described as follows:


\begin{itemize}
    \item \textbf{First Step (\textit{Initial Step with Finetuning}):} In the initial step the goal is to detect the most easily identifiable false positive pairwise predictions. We do this through selecting and labeling records from the components with the biggest numbers of false positive transitive predictions since they indicate the most flagrant violations of the Transitive Consistency. Additionally, we use the labeled pairs to further finetune $f_\theta$.
    \item \textbf{Second Step (\textit{Post Finetuning Cleanup \& Checks}):} The second step aims to remove from the matching as many false positives as possible. We do this by removing records from all components that violate the Transitive Consistency through a series of checks (which involve labeling) and a removal heuristic (which does not label removed pairs).
    \item \textbf{Final Step (\textit{Edge Recovery}):} The final step considers adding back some of the removed pairs (edges) while maintaining the Transitive Consistency of the resulting matching. This is done by evaluating the transitive matches that would be created if removed edges were to be added back to the matching.
\end{itemize}

TransClean is designed to operate within a predefined labeling budget $LB^{Total}$. The labeling can be done either manually or using pseudo-labels. In the experiments we will illustrate the performance difference between using an LLM to label pairs and manually labeling them. 

We now give detailed descriptions with pseudo-code for each of the steps as well as a visual example of the entire cleanup procedure (see Figure \ref{fig:SingleComponentViz} for a preview of how TransClean would process the component of Figure \ref{fig:Example_component_fig}).


\subsubsection{Initial Steps with Finetuning}\hfill\\
In the initial stages of TransClean, our primary goal is to detect the false positives that have the biggest effect in terms of wrongly linking different record groups in $G_{f_\theta}$ (such as the pairs ($r_{\#2}$, $r_{\#9}$)/($r_{\#4}$, $r_{\#9}$) and ($r_{\#3}$, $r_{\#6}$)/($r_{\#4}$, $r_{\#6}$) of Figure \ref{fig:Example_component_fig}). Detecting such matches and labeling them will also allow us to further fine-tune the pairwise matching model, potentially making it able to detect similar mistakes in other record groups.

In order to efficiently perform the evaluation over the whole graph of potential matches, we sort the components of $G_{f_\theta}$  by their number of negative transitive predictions. Next, we select from each component:

\begin{itemize}
    \item \textbf{Edges belonging to a \textit{Minimum Edge Cut}}. These are the edges that make up a set of minimum cardinality that if removed would disconnect the component. Minimum Edge Cuts of components with large numbers of negatively predicted transitive edges are likely to contain false positives. 
    
    To prevent the calculation of the Minimum Edge Cuts of exceedingly large components, we initially remove from the matching all subgraphs bigger than a threshold \textit{S}. 
    
    \item \textbf{Edges constituting \textit{shortest paths} between a subset of the records belonging to negatively predicted transitive matches in the component}. If a transitive edge is predicted negatively then it is likely to be a true negative pair. Since there cannot be an edge path of only true positives connecting the two records of a true negative pair,  selecting edges in this manner will likely yield false positives.
\end{itemize}

We carry out this edge selection and labeling process, removing from $G_{f_\theta}$ the edges labeled as false positives, until we reach a given labeling budget. Once the budget has been reached, we fine-tune $f_\theta$ with the edges that have been labeled (either manually or with pseudo-labels).

Additionally, we want to start leveraging the Transitive Consistency of $f_\theta$ in order to detect the inconsistent components of $G_{f_\theta}$ and break them up. To do this, we evaluate all the current transitive matches of $G_{f_\theta}$\footnote{If a component has more nodes than the threshold $S$, we only evaluate a subset of its transitive matches to avoid long running times, as the number of transitive matches grows quadratically with the number of nodes of the component.} with the fine-tuned model $f_{\theta^{TC}}$ and remove all the edges belonging to the Minimum Edge Cuts of components with more negative than positive transitive predictions. This pruning logic naturally decreases the number of negative transitive predictions, since the removal of Minimum Edge Cuts disconnects the sets of nodes which previously were creating the transitive matches. We expect to remove significantly more false positives than true positives with this logic, but the removal of some true positives is the price to pay for not labeling all the pairs considered this way. We repeat all of these steps $n$ times
, accumulating the newly labeled pairs at every step.

\begin{figure}[h!t]
  \centering
  \includegraphics[width=0.9\linewidth, keepaspectratio]{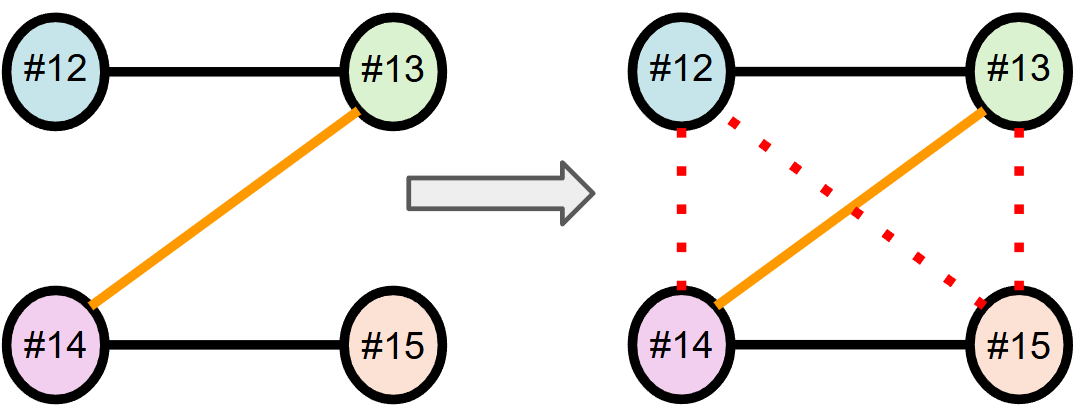}
  \caption{The component in this figure has two true positive matches (black edges) and a single false positive match between records $r_{\#13}$ and $r_{\#14}$ (orange edge). If any of the transitive edges ($r_{\#12}$, $r_{\#14}$), ($r_{\#12}$, $r_{\#15}$) and ($r_{\#13}$, $r_{\#15}$) (red dotting edges) is predicted as a $\texttt{NoMatch}$, then the model will not be transitively consistent in the component. In this case, all edges may be selected during Algorithm \ref{alg:Initial TransClean Step with Finetuning}. Either for fine-tuning, as a Minimum Edge Cut or part of a shortest path, or to be pruned without being labeled, also as a Minimum Edge Cut.}
  \label{fig:Edge Selection Viz}
\end{figure}

See Algorithm \ref{alg:Initial TransClean Step with Finetuning} for the pseudo-code description of this 1st step and Figure \ref{fig:Edge Selection Viz} for an illustration of the edge selection process. All successive steps carry out the same steps but starting out with  the fine-tuned model $f_{\theta^{TC}}$, the modified graph $G^{TC}$, and the labeled pairs $FT_{pairs}$ instead.

\begin{algorithm}
\caption{1st Initial TransClean Step with Finetuning}
\label{alg:Initial TransClean Step with Finetuning}
\begin{flushleft}
\textbf{Input:} Pairwise matching model $f_\theta$, matching graph $G_{f_\theta} = (V,E_{f_\theta})$, labeling budget $LB^{init}$, component size threshold $S$\\

\textbf{Output:} Finetuned model $f_{\theta^{TC}}$, modified $G^{TC}$
\end{flushleft}
\begin{algorithmic}[1]
\State $C = \{c_1, c_2, ..., c_n\}$  components of $G_{f_\theta}$ sorted by size
\State $C \gets LargeSubgraphBreakdown(C, S)$
\State $G^{TC} \gets C$ $\#$\textit{Initialize the graph}
\State $f_{\theta^{TC}} \gets f_{\theta}$ $\#$\textit{Initialize the matching model}
\State $LB_{spent} \gets 0$ $\#$\textit{Initialize the labeling budget}
\State $FT_{pairs} \gets \emptyset$ $\#$\textit{Initialize the set of finetuning pairs}
\State $i \gets 0$
\While{$ LB_{spent} < LB^{init}$}:
\State  $E_{\text{mincut}} \gets \text{MinEdgeCut($c_i$)}$ 
\State  $E_{RandomPaths} \gets \text{RandomPaths($c_i$)}$ 
\State $FT_{pairs} \gets FT_{pairs}  \bigcup E_{\text{mincut}}  \bigcup E_{RandomPaths}$ $\#$\textit{Accumulate the finetuning pairs}
\State $G^{TC} \gets Label(G^{TC}, FT_{pairs})$ 
\State $LB_{spent} \gets LB_{spent} + |E_{\text{mincut}}  \bigcup E_{RandomPaths}|$
\State $i \gets i+1$
\EndWhile

\State $f_{\theta^{TC}} \gets FineTune(f_{\theta^{TC}}, FT_{pairs})$
\State $G^{TC} \gets Prune_{MinEdgeCut}(G^{TC}, f_{\theta^{TC}})$ $\#$ \textit{Prune edges of the components with negatively predicted transitive matches}

\State \textbf{Return:} $f_{\theta^{TC}}, G^{TC}$
\end{algorithmic}
\end{algorithm}

\subsubsection{Post Finetuning Cleanup \& Checks}\hfill\\
After completing the initial TransClean stages, we expect to have removed most of the easily detectable false positives of the matching. At this point we have a model $f_{\theta^{TC}}$ that has been repeatedly fine-tuned and a modified graph $G^{TC}$ which has had the labeled false positive edges removed. Even though we have repeatedly removed Minimum Edge Cuts from the components with more negative transitive predictions than positive ones, there might still be components that fulfill this criterion, i.e. components with more negative transitive predictions than positive ones. To make sure we break up these components, we iteratively remove from $G^{TC}$ the Minimum Edge Cuts of components fulfilling the criterion until none that do remain.

At this point, we want to exploit the Transitive Consistency to its fullest in order to find the remaining false positives of $G^{TC}$. First, we check and label all the edges belonging to components with more nodes than the threshold $S$ mentioned in the previous section. Following the size check, we check whether any of the current transitive matches has been at any point labeled as a false positive. If any such match is currently implied by the remaining edges in $G^{TC}$, then there will be at least one false positive among the edges of its correspondent component, so we check and label all edges of components implying such matches and remove any labeled false positive edge.

Finally, in order to impose the Transitive Consistency of $f_{\theta^{TC}}$ in the current graph, we evaluate all remaining transitive matches and label all edges of components with any negative transitive predictions. We expect the labeling requirement in this step $LB^{{PostFT}}$ to be manageable, due to the extensive pruning we have previously performed in the initial steps. The amount of pairs left to check will be dependent on $f_{\theta^{TC}}$ and the current state of $G^{TC}$, unlike the labeling requirement of the initial steps which can be chosen instead. See Algorithm \ref{alg: Post Finetuning Cleanup & Checks} for a pseudo-code description of this stage.\\

\begin{algorithm}
\caption{Post Finetuning Cleanup $\&$ Checks}
\label{alg: Post Finetuning Cleanup & Checks}
\begin{flushleft}
\textbf{Input:} Pairwise matching model $f_{\theta^{TC}}$, matching graph $G^{TC}$, component size threshold $S$

\textbf{Output:} Modified $G^{TC}$, labeling requirement $LB^{{PostFT}}$
\end{flushleft}
\begin{algorithmic}[1]
\State $C \gets \{c_1, c_2, ..., c_n\}$  components of $G^{TC}$
\State $Stop_{flag} \gets False$
\While{NOT $Stop_{flag}$}: $\#$ \textit{Keep on pruning edges until the positive transitive predictions outnumber the negatives}
    \State $Stop_{flag} \gets True$
    \For{$c_i \in C$}:
        \State $Pos_{tr}(c_i) \gets \{(r_{i*}, r_{j*}) \in c_i^{trans} | f_{\theta^{TC}}((r_{i*}, r_{j*})) = \texttt{Match}\}$
        \State $Neg_{tr}(c_i)\gets\{(r_{i*}, r_{j*}) \in c_i^{trans} | f_{\theta^{TC}}((r_{i*}, r_{j*})) = \texttt{NoMatch}\}$
        \If{$|Pos_{tr}(c_i)| > |Neg_{tr}(c_i)|$}
            \State $G^{TC} \gets Prune_{MinEdgeCut}(c_i)$
            \State $Stop_{flag} \gets False$
        \EndIf
    \EndFor
    \State \textbf{Update} $C \gets \{c_1, c_2, ..., c_n\} $  components of $G^{TC}$

    \EndWhile
\State $G^{TC}, LB^{SC}  \gets SizeCheck(G^{TC}, S)$ 
\State $G^{TC}, LB^{LTC} \gets LabeledTransCheck(G^{TC})$ $\#$ \textit{Current transitive matches check}
\State $LB^{FinalTransCheck} \gets 0$ $\#$ \textit{Initialize the final check labeling effort}
\For{$c_i \in C$}: $\#$ \textit{Check all components with remaining negative transitive predictions}
    \State $Neg_{tr}(c_i)=\{(r_{i*}, r_{j*}) \in c_i^{trans} | f_{\theta^{TC}}((r_{i*}, r_{j*})) = \texttt{NoMatch}\}$
    \If{$|Neg_{tr}(c_i)| > 0$}
            \State $G^{TC}, LB^{MC} \gets Label(c_i)$ $\#$ \textit{Update the graph based on assigned labels and record size of checked component}
            \State $LB^{FinalTransCheck} \gets LB^{FinalTransCheck} + LB^{MC}$
        \EndIf
\EndFor
\State $LB^{{PostFT}} \gets LB^{SC} + LB^{LTC} + LB^{FinalTransCheck}$
\State \textbf{Return:} $G^{TC}$, $LB^{{PostFT}}$
\end{algorithmic}
\end{algorithm}

\subsubsection{Edge Recovery}\hfill\\
After carrying out all the Post Finetuning Cleanup $\&$ Checks we are left with a matching $G^{TC}$ in which $f_{\theta^{TC}}$ is transitively consistent\footnote{With the exception of pairs whose predictions might have changed due to the finetuning of $f_{\theta^{TC}}$ (but whose transitive matches are all predicted as positives) and components that have been fully labeled. We could choose to remove all pairs whose predictions become \texttt{NoMatch} at any point during the finetuning process, but this will likely remove numerous true positive pairs.}. To achieve this, we have removed edges from our original matching $G_{f_\theta}$ in multiple instances. We expect most of these matches to be correctly removed false positives, but some might be true positives, specially those removed during the initial stages, since the initial focus is to break up large components.

\begin{figure}[h!t]
  \centering
  \includegraphics[width=0.4\linewidth, keepaspectratio]{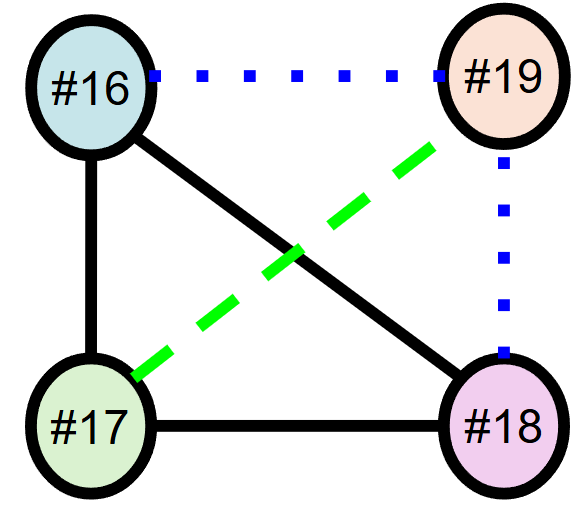}
  \caption{If we consider to recover the edge between records $r_{\#17}$ and $r_{\#19}$ (predicted as a match by $f_{\theta^{TC}}$), we will evaluate as new transitive matches the record pairs ($r_{\#16}$, $r_{\#19}$) and ($r_{\#18}$, $r_{\#19}$). If both are predicted as matches then we add the edge ($r_{\#17}$, $r_{\#19}$) to the matching, enlarging the component. If instead at least one of the record pairs is predicted as a \texttt{NoMatch}, we will label the edge ($r_{\#17}$, $r_{\#19}$) if the labeling budget allows for it, otherwise we will ignore it.}
  \label{fig:Edge Recovery Viz}
\end{figure}

We can once again leverage the Transitive Consistency of $f_{\theta^{TC}}$ in order to recover some of the true positive matches we have removed. We do this by evaluating the additional transitive matches that $G^{TC}$ would present if we added back each of the removed unlabeled edges that $f_{\theta^{TC}}$ currently predicts as a $\texttt{Match}$. If all of the transitive matches that the removed edge would create  if added back are predicted as a $\texttt{Match}$, then adding the removed edge would preserve the Transitive Consistency of $f_{\theta^{TC}}$ in $G^{TC}$, so we indeed add it back. If any of the transitive matches are predicted as a $\texttt{NoMatch}$ then we manually check the removed pair if the remaining labeling budget $LB^{EdgeRecovery} = LB^{Total} - LB^{init} - LB^{{PostFT}}$ allows for it, otherwise we ignore it. If the removed edge connects two records with no other matches, it will not create any new transitive match in $G^{TC}$ but adding it back would still respect the Transitive Consistency of $f_{\theta^{TC}}$ in $G^{TC}$, so we choose to label these records as well. Repeating this logic for every removed edge, we can add back to the matching difficult matches. See Algorithm \ref{alg: Edge Recovery} for a pseudo-code description of the process and Figure \ref{fig:Edge Recovery Viz} for a visual illustration of an instance of the edge recovery process.

Algorithm \ref{alg: Edge Recovery} requires separate evaluations of different sets of transitive matches with the pairwise model $f_{\theta^{TC}}$. All of the potential transitive matches we may need to evaluate can be determined at the beginning of the process, since they are the transitive matches of the components resulting from adding back to the matching \textbf{all deleted edges}. For speed considerations all of these evaluations can be carried out in batches, which is the setup most ML hardware is optimized for, but doing so requires carefully considering the size of the considered connected components. In our actual implementation, we use the component size threshold S, used in previous steps, to only evaluate the transitive matches of components smaller than S. We manually check the deleted edges that would lead to bigger components if the labeling budget $LB^{EdgeRecovery}$ allows for it or ignore them otherwise.

Note that adding a true positive edge to $G^{TC}$ may lead to an increase in the number of negative transitive predictions of $f_{\theta^{TC}}$ in $G^{TC}$ because the model may wrongly predict true positive transitive matches, as is particularly likely in record groups with edge cases. Negative transitive predictions can be ignored as long as they are caused by manually labeled matches, as the manual inspection constitutes stronger evidence than model predictions. It is only in unchecked components where the sign of the transitive predictions provides us with additional evidence, beyond the  predictions of the pairwise matching model $f_{\theta^{TC}}$, of the correctness of the component.

Finally, note that TransClean may not consume all of the initially predefined labeling budget $LB^{Total}$ if the amount of unlabeled removed edges predicted as a \texttt{Match} that create new negative transitive predictions is smaller than $LB^{EdgeRecovery}$. This is most likely when combining TransClean with robust pairwise matching models that easily reach Transitive Consistency on the initial stages of TransClean and do not flip their predictions on many removed unlabeled candidate edges after being finetuned.

\begin{algorithm}
\caption{Edge Recovery}
\label{alg: Edge Recovery}
\begin{flushleft}
\textbf{Input:} Pairwise matching model $f_{\theta^{TC}}$, matching graph $G^{TC}$, removed matches $\mathcal{R} = \{(r_i, r_j)_{rem} | f_{\theta^{TC}}((r_i, r_j)_{rem}) = \texttt{Match}\}$, labeling budget $LB^{EdgeRecovery}$

\textbf{Output:} Final matching $G^{TC}$
\end{flushleft}
\begin{algorithmic}[1]
\For{$(r_i, r_j)_{rem} \in \mathcal{R}$} $\#$ \textit{Check for each removed edge, which component it would belong to/create if added back}
\State $c_i \gets getComponent(r_i, G^{TC})$ $\#$ 
\State $c_j \gets getComponent(r_j, G^{TC})$ $\#$ 
    \If{$c_i == c_j$}
        \State \textbf{Continue} $\#$ \textit{This edge is already in the matching (as a transitive match) since both records already belong to the same component.}
    \Else{}
    \State $Preds \gets EvalNewTransMatches((r_i, r_j), G^{TC}, f_{\theta^{TC}})$

        \If{$Preds = \emptyset$ and $LB^{EdgeRecovery}$ > 0}

        \State $\#$ \textit{Edge connects only a pair of records.}
        \State $G^{TC} \gets ManualCheck((r_i, r_j), G^{TC})$
        \State $LB^{EdgeRecovery} \gets LB^{EdgeRecovery} - 1$
        
        \ElsIf{$pred == \texttt{Match}$ $\forall$  $pred \in Preds$}:
            \State $\#$ \textit{This edge connects two different components into a transitively consistent component.}
            \State $G^{TC} \gets AddBack((r_i, r_j), G^{TC})$
        \ElsIf{$LB^{EdgeRecovery}$ > 0}:
            \State $\#$ \textit{This edge produces a connected component that is not transitively consistent.}
            \State $G^{TC} \gets ManualCheck((r_i, r_j), G^{TC})$
            \State $LB^{EdgeRecovery} \gets LB^{EdgeRecovery} - 1$
        \EndIf
    \EndIf
\EndFor

\State \textbf{Return:} $G^{TC}$
\end{algorithmic}
\end{algorithm}

\subsubsection{LLM Labeling}\label{Section:LLMLabeling}\hfill\\
When running TransClean, several steps of Algorithms \ref{alg:Initial TransClean Step with Finetuning}, \ref{alg: Post Finetuning Cleanup & Checks} and \ref{alg: Edge Recovery} involve the labeling of record pairs. This can be done either manually or using the pseudo-labels produced by a different model than the one used to produce the matching. In the experiments, we use an LLM for the labeling involved in the Initial TransClean Steps with Finetuning and the Post Finetuning Cleanup \& Checks (Algorithms \ref{alg:Initial TransClean Step with Finetuning}, \ref{alg: Post Finetuning Cleanup & Checks}) but not in the Edge Recovery step (Algorithm \ref{alg: Edge Recovery}) due to the slower inference time of LLMs and the large number of record pairs requiring labels in this final step.

A drop in performance can be expected from using pseudo-labels since some true positives may be wrongly labeled as non matches and equivalently with false positives being labeled as true matches. In the experiments we compare the performance of TransClean using LLM pseudo-labels vs manual labeling and illustrate its robustness in effectively detecting false positives even when the LLM produces the wrong labels.

\subsubsection{TransClean Visual Example}
\begin{figure}[h!t]
  \centering
  \includegraphics[width=\linewidth, keepaspectratio]{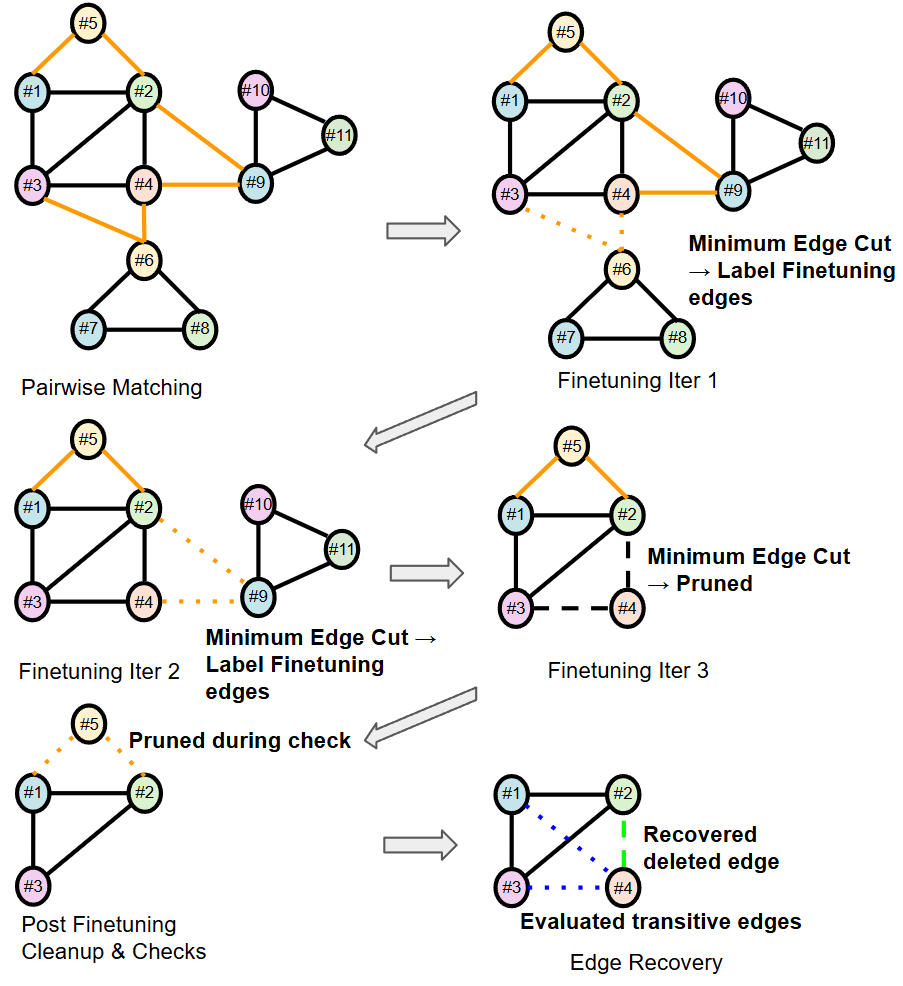}
  \caption{Evolution of the component of Figure \ref{fig:Example_component_fig} over the different steps of TransClean. Black edges illustrate true positive edges, orange edges false positive edges and dotted edges denote the selected edges, either for labeling or pruning, during a specific TransClean step.}
  \label{fig:SingleComponentViz}
\end{figure}

\hfill\\
Figure \ref{fig:SingleComponentViz} provides a visualization of all the steps of TransClean involved in cleaning up the component presented in Figure \ref{fig:Example_component_fig}. TransClean starts off by applying three \textit{Initial Steps with Finetuning} with the goal to detect false positive edges via Minimum Edge Cuts and shortest paths (see edges ($r_{\#3}$, $r_{\#6}$), ($r_{\#4}$, $r_{\#6}$) in Finetuning Iter 1 and edges ($r_{\#2}$, $r_{\#9}$), ($r_{\#4}$, $r_{\#9}$)  in Finetuning Iter 2 with orange dotted lines) to further finetune the pairwise matching model. Each of these edges is labeled, consuming the labeling budget, and consequently removed from the component (since they are false positives). Additionally, the Finetuning Iter 3 pruning step, which is done without labeling,  removes 2 true positive edges  (see the edges $r_{\#2}$, $r_{\#4}$) and ($r_{\#3}$, $r_{\#4}$)), since they form a minimum edge cut of the component. This is an example of removing a true positive match which needs to be recovered in the final edge recovery stage. These steps iteratively break up the component.

The last false positive edges ($r_{\#1}$, $r_{\#5}$) and ($r_{\#2}$, $r_{\#5}$) are pruned from the component during the \textit{Post Finetuning Cleanup and Checks} and the removed true positive edges $\{(r_{\#i}, r_{\#4})\}_{i=1}^3$ are recovered during the \textit{Edge Recovery} step. 

\section{Experiments}\label{Section:Experiments}
In this section we perform several end-to-end matching experiments with TransClean to address the following research questions:

\begin{itemize}
    \item \textit{How can TransClean be leveraged to provide an estimation of the quality of a given matching?}
    \item \textit{Is TransClean able to improve matchings by detecting most of their false positive matches?}
    \item \textit{How dependent is the performance of TransClean on the quality of the pairwise matching model it is combined with?}
    \item \textit{What is the performance gap between manually labeling the candidate pairs selected by TransClean and pseudo-labelling them with an LLM?}
\end{itemize}
All our experiments were run on an Ubuntu machine with a single Nvidia Tesla T4 GPU, 16 VCPUs, and 64 GB of RAM.

\begin{table*}[]
\caption{General dataset statistics. Highlighted in bold are the number of records we use during model training and the matching experiments.}
\label{tab:dataset_statistics}

\begin{tabular}{l|r|r|r|r|r}
\thickhline \textbf{Dataset} & Synthetic Companies & MusicBrainz & Camera & Monitor  & WDC Products\\
\thickhline  \textbf{\# of data sources} & 5 & 5 & 24 & 26 & 3,259\\
\textbf{\# of records} & 868K & 19K & 3.9K & 2.3K & 4.8K\\

\textbf{\# of record groups} & 200K & 10K & 103 & 252 & 2.2K\\
\textbf{\# of matches} & 1.5M  & 16.25K & 157K & 13.5K &  28.3K\\
\begin{tabular}{@{}l@{}}\textbf{\# of records in} \\ \texttt{train/val/test}\end{tabular} & 520K\textbf{(4K)}/\textbf{174K}/\textbf{174K} & \textbf{14K/1.8K/456} & \textbf{2.5K/696/632} & \textbf{1.5K/570/196} & \textbf{2.8K}/\textbf{1K}/\textbf{1K}\\
\textbf{Attributes} & \begin{tabular}{@{}c@{}}\textit{name, city, region,} \\ \textit{country\textunderscore code,} \\ \textit{short\textunderscore description}\end{tabular} & \begin{tabular}{@{}c@{}}\textit{number, length, artist, } \\ \textit{year, language}\end{tabular} & \begin{tabular}{@{}c@{}}\textit{multiple} \\ \textit{(.json format)}\end{tabular} & \begin{tabular}{@{}c@{}}\textit{multiple} \\ \textit{(.json format)}\end{tabular}  & \begin{tabular}{@{}c@{}}\textit{brand, title, description,} \\ \textit{price, priceCurrency}\end{tabular} \\
\thickhline 
\end{tabular}
\end{table*}

\subsection{Experiments Setup}\label{Experiments:Experiments Setup}

\subsubsection{Datasets}\label{Experiments:Dataset}
We experiment with the following multi-source datasets:

\begin{itemize}
    \item \textbf{\textit{Synthetic Companies}}: Presented in \cite{DBLP:conf/edbt/PardoLGNNMBS25}, this is a procedurally generated benchmark explicitly designed to mimic real-world matching conditions containing 868K records To the best of our knowledge, this is currently the largest entity matching data set used in the literature. Fixed training, validation, and test splits are provided. Additionally, it includes  a set of 7K positive train pairs (involving 4K records) obtained via labeling the first 10K pairs through a heuristic blocking\footnote{We use this subset of labeled pairs during training, as being able to label all training pairs would be unrealistic in a real-world setting.}.

    \item \textbf{\textit{MusicBrainz}}: Presented in \cite{Saeedi2017}, this benchmark is based on song records from the MusicBrainz database and uses a data generator to create duplicates with modified attribute values. We experiment on the 20k variant\footnote{All the differently sized variants can be retrieved from \href{https://dbs.uni-leipzig.de/research/projects/benchmark-datasets-for-entity-resolution}{https://dbs.uni-leipzig.de/research/projects/benchmark-datasets-for-entity-resolution}}. 

    \item \textbf{\textit{Camera/Monitor}}: Presented in \cite{AlaskaDataBenchmark}, these benchmarks consist of web-scraped JSON-files, each with different properties, containing product information from 24 and 26 different websites respectively. For both datasets, we cast all records into a table format by recording the \texttt{page\textunderscore id} and merge all other attributes into a \texttt{text} attribute.

    \item \textbf{\textit{WDC Products}}: Presented in \cite{2023WDCProducts}, this benchmark consists on product data retrieved from  3,259 e-shops. Multiple variants of the dataset are available, varying in terms of the amount of corner-cases, unseen entities and development set size. We experiment on the variant with $80\%$ corner cases and a test set with $100\%$ unseen entities. Fixed training, validation, and test splits are provided.

\end{itemize}

See Table \ref{tab:dataset_statistics} for the general statistics of each of the datasets. For the MusicBrainz and Camera/Monitor datasets we implement our own train/val/test splits. We split based on a 60/20/20 proportion across the record groups, leading to differently sized splits (record groups have different numbers of records). This splitting makes sure that no test records are used during training. Unpaired records in the ground truth of the MusicBrainz dataset do not get assigned to any split.

\subsubsection{Machine Learning Models}\label{Experiments:MLModels}
We employ 2 different pairwise matching ML models in our experiments. DistilBERT \cite{2019DistilBERT}, a distilled version of BERT, and CLER \cite{CLER2024}, a larger\footnote{CLER uses RoBERTa as a classifier, which has 355M parameters compared to the 66M of DistilBERT.} state-of-the-art methodology explicitly developed for EM. We train both models as follows: 

\begin{itemize}
    \item \textbf{DistilBERT:} We fine-tune DistilBERT for 5 epochs and select the best performing epoch on the validation split. In case the dataset does not include negative pairs, we add them randomly in a proportion of 5 to 1.
    \item \textbf{CLER:} We train CLER with different labeling budgets\footnote{CLER trains via an uncertainty based active learning technique. As such, it requires an initial labeling budget to determine the number of training pairs to select. We use 10K for the Synthetic Companies, 5K for MusicBrainz \& Camera, 2K for Monitor and 8K for WDC.}. As validation sets (used in CLER's training logic) we choose 500 randomly selected train pairs, as done in CLER's original implementation.
\end{itemize}
We do not use Large Language Models (LLMs) for pairwise matching due to their comparatively slower inference times. Rather, we use \textit{DeepSeek LLM 7B Base} for labeling the candidate pairs selected by TransClean in the runs without manual labeling. We carry out this labeling by prompting the model with the following prompt: 

\begin{center}
    \texttt{"Do these two records represent the same entity? Answer only Yes or No, do not elaborate further. First record: [$record_1$] Second Record: [$record_2$]}
\end{center}
And recover the labels from the LLM output via a regex, where we consider an output to be a \texttt{Match} prediction if it contains the substring \textit{"yes"} and a \texttt{NoMatch} prediction otherwise.

\subsubsection{Blocking}\label{Experiments:Blocking}
Evaluating all possible record pairs in the test sets would be prohibitive\footnote{For the Synthetic Companies dataset the number of records in the test set is 174K so we would need to evaluate (174K) * (174K - 1)/2 = 1.5 *$10^{10}$ record pairs.} hence we need a blocking approach to reduce the amount of records that need to be evaluated. For DistilBERT we use a simple token overlap heuristic that pairs each record with the top 10 records ranked by the number of shared tokens between their token sequences. On the Synthetic Companies dataset we also use the token overlap heuristic used in \cite{DBLP:conf/edbt/PardoLGNNMBS25}. CLER on the other hand, produces both an embedding-based \texttt{blocker} and a pairwise \texttt{matcher}. We use the \texttt{matcher} to evaluate  the top 10 most similar pairs produced by the \texttt{blocker}.
See Table \ref{tab:trainlb_and_candidate_pairs} for a summary of the sets of candidate pairs used for each model along with their recall scores (the percentage of total true positives included in each set of pairs).

\begin{table}[H]
\caption{Number of test candidate pairs considered by each of the models and their associated recall score.}
\label{tab:trainlb_and_candidate_pairs}

\begin{tabular}{c|c|c }
\thickhline \textbf{Dataset} &\textbf{Model} &   \begin{tabular}{@{}c@{}}\# of test candidate\\  pairs $\&$ recall \end{tabular}  \\
\thickhline \begin{tabular}{@{}c@{}}\textbf{Synthetic} \\ \textbf{Companies}\end{tabular}  &\textbf{DistilBERT} & 1.44M - 96.10$\%$ \\

& \textbf{CLER}  & 430K - 98.25$\%$ \\
\thickhline 
\textbf{MusicBrainz} & \textbf{DistilBERT} & 6.5K - 99.34$\%$ \\
& \textbf{CLER} & N/A\tablefootnote{On the MusicBrainz dataset CLER's training failed to conclude after 40 hours of continuous training, so we omit it from our analysis.}  \\
\thickhline
\textbf{Camera} & \textbf{DistilBERT} & 5.9K - 83.38$\%$ \\
 & \textbf{CLER} & 2.1K - 68.74$\%$ \\
\thickhline
\textbf{Monitor} & \textbf{DistilBERT}  & 1.9K - 95.31$\%$ \\
 & \textbf{CLER} & 519 - 88.02$\%$ \\
\thickhline

\begin{tabular}{@{}c@{}}\textbf{WDC} \\ \textbf{Products}\end{tabular} & \textbf{DistilBERT} & 9.2K - 81.8$\%$\\
 & \textbf{CLER} & 3.6K - 62.6$\%$ \\
\thickhline

\end{tabular}
\end{table}

\subsubsection{TransClean}\label{Experiments:TransClean}\hfill\\

We run TransClean on the matchings produced by the pairwise predictions of both models. We set the number of initial steps with finetuning to $n = 5$ and the component size threshold to $S=50$. We allocate $\frac{LB^{Total}}{2}$ to the Initial TransClean steps with finetuning and the remaining budget to the following steps.

\subsection{Pairwise matching vs TransClean}\label{Section:PairwiseMatchingvsTransClean}

\subsubsection{Pairwise Matching}\hfill\\
We start off our discussion by highlighting the challenges associated with attempting to match the Synthetic Companies dataset through pairwise matching only. A standard attempt would start by training CLER with a 10K labeling budget and the records involved in the labeled pairs described in Section \ref{Experiments:Dataset}. This will produce 2 models, a \texttt{blocker} and a \texttt{matcher}. The \texttt{blocker} produces a set of 430K candidate pairs, as noted in Table \ref{tab:trainlb_and_candidate_pairs}, of which the \texttt{matcher} considers 213K as matches. The question now is how to evaluate the quality of the produced matching. Blindly trusting the predictions of the \texttt{matcher} is risky, due to the likelihood of false positive matches. On the other extreme, manually checking all of the produced matches would require an impossibly large human effort. It is then natural to consider the transitive matches, which are implied by the pairwise predictions but haven not been evaluated by the model yet.

\subsubsection{Transitive Matches}
\begin{figure}[h!t]
  \centering
  \includegraphics[width=\linewidth, keepaspectratio]{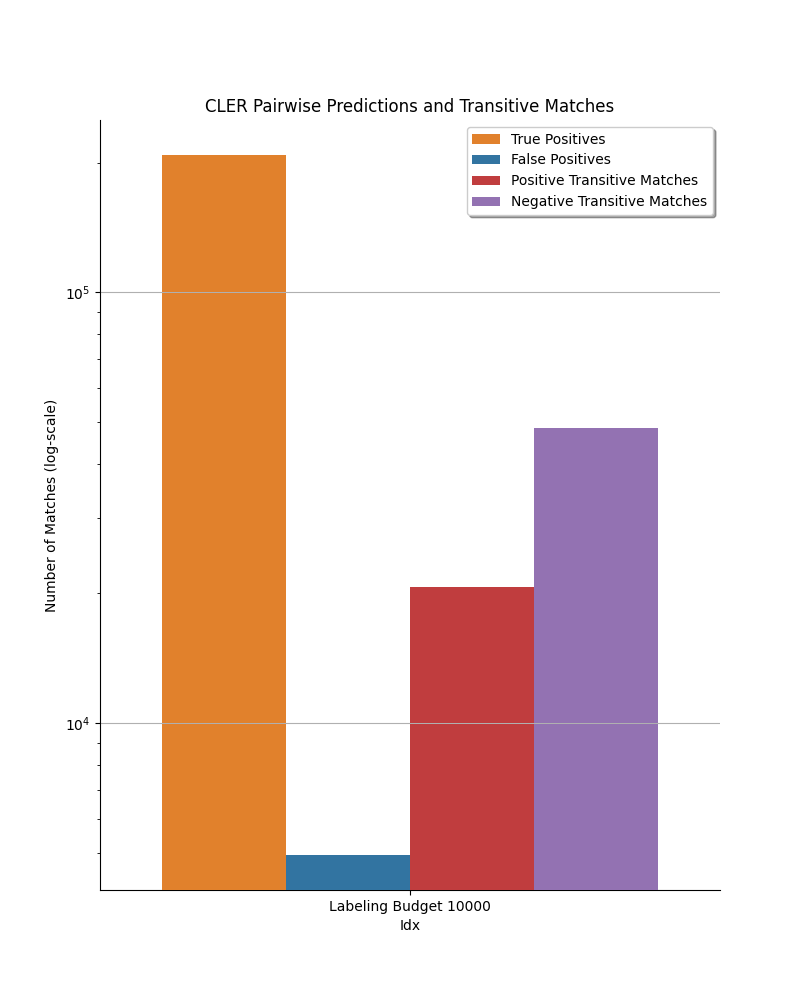}
  \caption{Aggregate pairwise predictions on the Synthetic Companies dataset produced by CLER trained with a labeling budget of 10K pairs. Even though the \texttt{matcher} model has a high precision, the number of negative transitive matches indicates a lack of Transitive Consistency and thus the likely presence of false positive matches.}
  \label{fig:PairwiseVizTransPreds}
\end{figure}

\hfill\\
As depicted in Figure \ref{fig:PairwiseVizTransPreds}, the 213K predicted matching pairs of CLER lead to 69K transitive matches of which 20.5K are predicted as matches and 48.5K as non-matches. This high number of negatively predicted transitive matches indicates that the \texttt{matcher} is not transitively consistent and that our matching likely has a considerable number of false positive matches.

\begin{figure*}[h!t]
  \centering
  \includegraphics[width=1.0\textwidth, keepaspectratio]{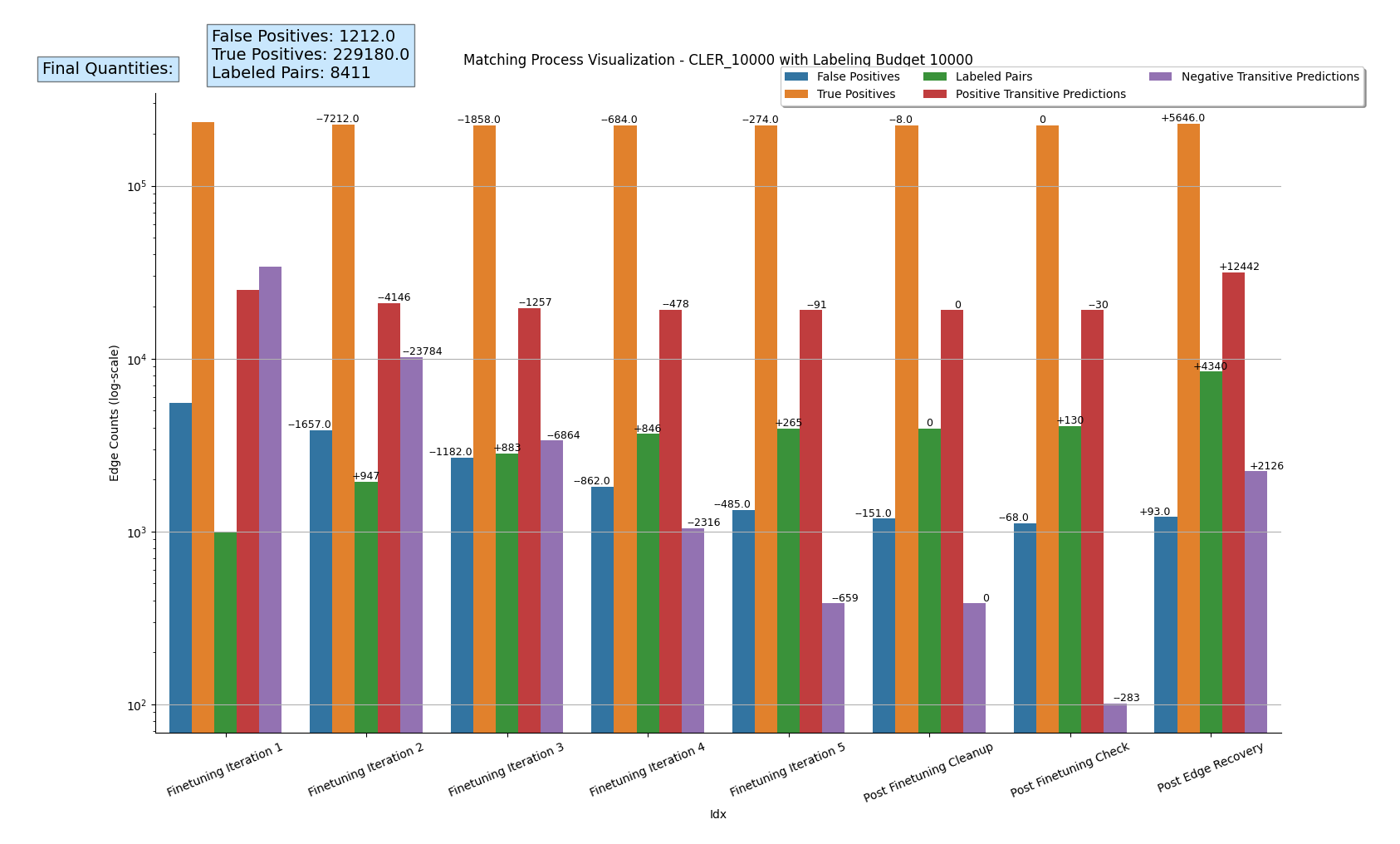}
  \caption{Visualization of relevant metrics of a matching on the Synthetic Companies dataset and their evolution over the different steps of TransClean.  Each set of bars illustrates from left to right the number of: 1. False Positives 2. True Positives 3. Labeled Pairs 4. Positive Transitive Predictions 5. Negative Transitive Predictions with their variations w.r.t to the previous column from the second Finetuning Iteration 2 onward. The amounts of true and false positives include all the matches present in the matching, both predicted and transitive.}
  \label{fig:TransCleanviz}
\end{figure*}

\subsubsection{Running TransClean}\hfill\\
In order to efficiently find the false positive matches, we execute TransClean with a 10K labelling budget as specified in Section \ref{Experiments:TransClean} on the matching produced by the 213K  predicted matching pairs. Figure \ref{fig:TransCleanviz} illustrates the evolution of the most relevant metrics of the matching during the different TransClean stages. 

We start off labeling 1K record pairs in the first Finetuning Iteration and iteratively label up to 10K pairs through the different TransClean steps (see green bars). Notice how the number of negative transitive predictions (see purple bars) steadily decreases\footnote{Except for the last Post Edge Recovery step, where the manually labeled pairs add transitive pairs that are true matches but are incorrectly predicted by the \texttt{matcher}.} whereas the number of positive transitive predictions (see red bars) largely remains constant. These two quantities indicate the gradual success of TransClean in progressively modifying the matching towards being transitively consistent while not removing too many of the initial true matches. 

Note also the \textit{correlations of both quantities with the amounts of true and false positives}, which makes them an ideal pair of proxies. In other words, positive transitive predictions are correlated with true positives, and negative transitive predictions are correlated with false positives. While the amounts of true and false positives are unobtainable in practice (since they would require manually labeling the entire matching), the numbers of positive and negative transitive predictions can be easily obtained, and are in fact calculated at each step of TransClean, simply by evaluating the transitive matches.

\begin{table*}[]

\caption{Scores achieved by each combination of pairwise classifier and TransClean settings (labeling budget vs LLM labeling). Ditto marks indicate the repetition of the value above in the respective column (experiments that use the same pairwise matching model will have the same Pairwise Matching and Pre TransClean scores and Training/Inference times). Note that the labeling budget and the LLM generated labeling (see column "TransClean Variant") are only used during the TransClean phase.}
\label{tab:group_matching_table}
\resizebox{\textwidth}{!}{%
\begin{tabular}{c|c|c|ccc|ccc|ccccc|ccc}

 & \begin{tabular}{@{}c@{}}\textbf{Pairwise} \\ \textbf{Matching} \\ \textbf{Model}  \end{tabular}
 & {\begin{tabular}{@{}c@{}}\textbf{TransClean} \\ \textbf{Variant} \end{tabular}}  
 &\multicolumn{3}{c|}{\begin{tabular}{@{}c@{}}\textbf{Pairwise Matching} \\ (pairs from blocking)\end{tabular}} 
 &  \multicolumn{3}{c|}{\begin{tabular}{@{}c@{}}\textbf{Pre TransClean} \\ (including transitive matches) \end{tabular}} 
 &  \multicolumn{5}{c|}{\begin{tabular}{@{}c@{}}\textbf{Post TransClean} \\ (including transitive matches) \end{tabular}}  & \multicolumn{3}{c}{\begin{tabular}{@{}c@{}}\textbf{Running Times} \\ hh:mm:ss  \end{tabular}}\\
\textbf{Dataset} & & & \textbf{Precision} & \textbf{Recall} & \textbf{F1 Score}  &  \textbf{Precision} & \textbf{Recall} & \textbf{F1 Score}  & \begin{tabular}{@{}c@{}}$\%$ \\ \textbf{True Positives} \\ \textbf{Removed} \end{tabular}
& \begin{tabular}{@{}c@{}}$\%$ \\ \textbf{False Positives} \\ \textbf{Removed} \end{tabular}&  \textbf{Precision} & \textbf{Recall} & \textbf{F1 Score} & \begin{tabular}{@{}c@{}}\textbf{Training} \\ Time \end{tabular} & \begin{tabular}{@{}c@{}}\textbf{Inference} \\ Time \end{tabular} & \begin{tabular}{@{}c@{}}\textbf{TransClean} \\ Time \end{tabular} \\ \hline
  \begin{tabular}{@{}c@{}}\textbf{Synthetic} \\ \textbf{Companies}  \end{tabular}                & \textbf{DistilBERT} & 10K budget    & 85.86   & 77.63  & 81.54  & 0.02 & 83.68 & 0.04 & 36.35$\%$ & 96.07$\%$ & 94.7 & 54.85 & 69.47 & 11:32:20 & 01:09:24  & 02:36:44 \\
                            & \textbf{DistilBERT} & \begin{tabular}{@{}c@{}}LLM \\ Labeling \end{tabular}  & "  & "  & "  & "  & " & "   & 42.00$\%$  & 96.67$\%$  & 97.61 & 47.71 & 64.09 & " &" & 27:51:32   \\
                          & \textbf{CLER}  & 10K budget  & 97.69  & 77.79  & 86.61 & 87.48 & 86.87 & 87.17 & 1.88$\%$ & 78.06$\%$  & 98.54 & 86.67 & 92.22 & 14:49:57 & 01:36:37 &  01:18:42  \\
                          
                           & \textbf{CLER}   & \begin{tabular}{@{}c@{}}LLM \\ Labeling \end{tabular}  & "  & "  & "  & "  &" & " & 3.52$\%$ & 72.06$\%$  & 99.02 & 82.83 & 90.2 & "  &" & 18:00:15   \\

                          \hline
  \begin{tabular}{@{}c@{}}\textbf{MusicBrainz} \\  \end{tabular}                & \textbf{DistilBERT} &1K budget  & 94.74  & 98.68  & 96.67  & 86.04 & 98.68 & 91.93  & 0$\%$ & 56$\%$ & 97.61 & 98.68 &  98.14 & 01:16:16 & 00:00:26 & 00:00:12 \\
                            & \textbf{DistilBERT} & \begin{tabular}{@{}c@{}}LLM \\ Labeling \end{tabular}  & "  & "  & "  & "  & " & "   & 3.11$\%$  & 48.0$\%$  & 96.9 & 95.83 & 96.36 & "  &" & 00:09:15  \\
                          & \textbf{CLER}  & \textit{N/A}  & -- & -- & -- & -- & -- & -- & -- & --  & -- & -- & -- & \textit{Out of time} & -- & --        \\
                          \hline
  \begin{tabular}{@{}c@{}}\textbf{Camera} \\  \end{tabular}                & \textbf{DistilBERT} & 1K budget    & 86.00  & 2.44  & 4.75  & 66.88 & 6.49 & 11.83 & 0$\%$ & 76.19$\%$ & 96.8 & 6.49 & 12.16 & 11:28:32 & 00:00:55 & 00:00:32 \\
                            & \textbf{DistilBERT} & \begin{tabular}{@{}c@{}}LLM \\ Labeling \end{tabular}  & "  & "  & "  & "  & " & "   & 87.59$\%$  & 73.02$\%$  & 55.81 & 0.3 & 0.6 & "  &" &  00:27:17 \\
                          & \textbf{CLER}  &  1K budget  & 98.5  & 0.83 & 1.65 & 97.27 & 1.12 & 2.21 & 0$\%$ & 50$\%$  & 99.44 & 1.12 & 2.22 & 16:19:32 & 00:07:30 &  00:08:07 \\
                          & \textbf{CLER}  & \begin{tabular}{@{}c@{}}LLM \\ Labeling \end{tabular}  & " & " &  " & " & " & " & 35.88$\%$  & 50$\%$ & 98.84 & 0.54 & 1.07 & " & " & 00:10:31     \\

                          \hline
  \begin{tabular}{@{}c@{}} \textbf{Monitor} \\  \end{tabular}                & \textbf{DistilBERT} & 1K budget    & 36.12  & 23.79  & 28.69  & 37.98 & 38.35 & 38.16 & 8.16$\%$ & 57.31$\%$ & 60.19 & 31.55 & 41.4 & 00:55:39 & 00:00:10 & 00:00:55 \\
                            & \textbf{DistilBERT} & \begin{tabular}{@{}c@{}}LLM \\ Labeling \end{tabular}  & "  & "  & "  & "  & " & "   & 30.61$\%$  & 13.46$\%$  & 35.89 & 27.99 & 31.45 & "  &" & 00:38:42  \\
                          & \textbf{CLER}  & \textit{N/A}  & 100  & 14.72 & 25.66 & 100 & 14.72 & 25.66  & -- & --  & -- & -- & -- & 09:46:01 & 00:02:57 &    --     \\

                          \hline

  \begin{tabular}{@{}c@{}}\textbf{WDC} \\ \textbf{Products}  \end{tabular}                & \textbf{DistilBERT} &\begin{tabular}{@{}c@{}} 10K budget \\ (-- used) \end{tabular}    & 34.52  & 63.8  & 44.8  & 1.13  & 66.6 & 2.22  & 12.22$\%$ & 86.77$\%$ & 75.54 & 56.2 & 64.45 &  02:14:00 & 00:00:40 & 00:02:12  \\
                            & \textbf{DistilBERT} & \begin{tabular}{@{}c@{}}LLM \\ Labeling \end{tabular}  & "  & "  & "  & "  & " & "   &  46.7$\%$   & 77.52$\%$  & 49.86  & 34.6 & 40.85 & "  &" &  01:36:32  \\
                          & \textbf{CLER}  & \textit{N/A}  & 100  & 0.2  & 0.4  & 100  & 0.2 &  0.4 & -- & --  & -- & -- & -- & 14:39:32  & 00:10:10 & --        \\
                          \hline

\end{tabular}
}
\end{table*}

\subsection{Results for All Datasets}\label{Section:Results}

Table \ref{tab:group_matching_table} presents the scores achieved by running TransClean, with manual labeling and LLM pseudo-labels, on the matching produced by each pairwise matching model for all datasets. The pairwise matching scores reflect the scores achieved considering only pairwise predictions. However, the Pre \& Post TransClean scores incorporate the transitive matches, which makes them a better indicator of the quality of the matchings, as discussed in Section \ref{Section: Introduction}.
\subsubsection{Pairwise Matching Results}\hfill\\

We start our discussion with the \textit{Pairwise Matching} and \textit{Pre TransClean} scores. Both DistilBERT \& CLER achieve in the Synthetic Companies dataset high F1 scores for pairwise matching ranging from 81.54 to 86.61. However, notice that with the inclusion of transitive matches the  F1 score for DistilBERT drops to 0.04. This is expected, since transitive matches are true matches if they connect records from the same ground truth component (which increases the recall of the matching) but are non matches (i.e. counted as false positives and thus decrease the precision) otherwise. In the case of DistilBERT, a high pairwise F1 score of 81.54 hides the large number of erroneous transitive matches implied by the false positives the model produces\footnote{These false positives lead to far more and far larger connected components, which imply the large number of erroneous transitive matches.}, which make the \textit{Pre TransClean} precision drop to 0.02. For CLER the inclusion of transitive matches has a positive impact on the Pre TransClean F1 score, which increases from 86.61 to 87.17 since the decrease in precision caused by adding the transitive matches is counteracted by the increase in recall. These results indicate that CLER's matcher is more robust than DistilBERT.

In the MusicBrainz dataset, DistilBERT achieves again high precision and recall, this time with a much smaller drop in terms of precision and F1 scores when adding into the matching the transitive matches. CLER on the other hand fails to conclude its training process after $40h$ of continuous training.

In the Camera and Monitor datasets both pairwise matching models yield matchings with very low recall. This is surprising, given the close-to-perfect scores reported in \cite{Yao2022EntityRW} and \cite{CLER2024} but both of these works only experiment with a subset of the records (since no fixed splits are provided). \cite{Yao2022EntityRW} only compares records against those of the same category and \cite{CLER2024} uses only 2 data sources out of the 24 included in the monitor dataset. In turn, we use all labeled records from both datasets. The low F1 scores achieved by both models signify that the actual difficulty posed by these datasets has been largely sidestepped in previous works through down-sampling.

Finally in the WDC Products dataset, DistilBERT produces a matching with low pairwise precision and consequently also very low Pre TransClean precision. CLER on the other hand, produces a model with extremely low recall but perfect precision, similarly as in the Camera and Monitor datasets. This signifies that CLER's uncertainty based training is not successful in datasets with a lot of diversity and inherently difficult records as it produces models with very low recall. 

\subsubsection{TransClean Results} \hfill\\

TransClean improves the matchings produced by all pairwise models\footnote{Except for the CLER on the Musicbrainz dataset where it fails to conclude its training process and on the Monitor and WDC Products datasets since the matches it produces have perfect precision but extremely low recall. Since these matchings have no false positives, TransClean cannot be applied.}, with an average F1 score improvement of +24.42 between Pre and Post TransClean scores when run with manual labeling, illustrating the importance of removing false positives from matchings. Across all datasets, TransClean manages to consistently remove a substantial proportion of false positives from the matchings produced by both models. However, when run with DistilBERT it removes a much higher percentage of true positives than when run with CLER because the former produces a lot more false positives and thus needs to remove a lot more pairs from the matchings in order to reach Transitive Consistency. 

When combined with an effective pairwise matching model, as is the case of both models in the MusicBrainz dataset, DistilBERT in the WDC products and CLER in the Synthetic Companies, TransClean removes very few true positives, leading to a substantial improvement in F1 scores. This illustrates how TransClean can detect most of the false positives of well performing models. 

When using an LLM to produce pseudo-labels, we can observe that the percentage of removed true positives increases while that of the removed false positives largely decreases across all datasets. However the differences in the datasets where the pairwise matching models achieve good results (Synthetic Companies and MusicBrainz) are small, due to both the LLM performance and TransClean's tolerance to pseudo-labels. In the other datasets, labeling with the LLM leads to removing large proportions of the original true positives, both because the matchings start with very few matches due to the low recall and because in these datasets the transitive matches are not a good indication of the matching's quality due to the poor performance of the pairwise matching models.

In terms of running times, training DistilBERT as we describe in Section \ref{Experiments:MLModels} is faster than training CLER, whose uncertainty based active learning training leads to much longer training times. Inference times are similar for both models, given that they both have a relatively low number of parameters for today's standards, with CLER being slower than DistilBERT due to its \texttt{matcher} having 5 times more parameters. Finally, in terms of TransClean times, the matchings produced by DistilBERT have a lot more false positives and thus TransClean takes longer to run. Additionally, the runs with LLM labeling are slower due to the additional time it takes to do inference to produce pseudo-labels with the LLM. In the runs with manual labeling, the labeling is done with a ground truth lookup and thus is not reflected in the corresponding times, for a fair comparison between both setups, an additional labeling time should be considered.

\section{Conclusions}\label{Section: Conclusions and Future Work}

In this work, we have presented TransClean, a method that leverages the Transitive Consistency of a pairwise model in a given matching in order to gradually remove its false positive matches while removing few true matches. We have shown how the proposed method can be used to remove false positives from matches produced by pairwise matching methods trained naively (DistilBERT) as well as state-of-the-art matchers (CLER) and analyzed the differences in its performance when using an LLM to produce pseudo-labels compared to doing manual labeling. 

Beyond entity matching, the principles of Transitive Consistency may generalize to domains like knowledge graph refinement, author disambiguation, or anomaly detection, where consistency across relational structures is key. We plan to explore these applications in future work.

We have shown how the positive and negative transitive predictions calculated during the different stages of TransClean can be used as proxies for the true and false positives of a matching as well as leveraged to detect false positives if the pairwise matching model used to produce the matching is effective. Our experiments show that, in terms of the actual record groups produced in multi-source matching settings, the effect of even small quantities of false positives should not be ignored. However, TransClean is able to detect such model mistakes if it is combined with a well performing pairwise matching model.


\begin{acks}
We thank Andrea Nagy and Barnabas Gera, both formerly of Move Digital AG, for their valuable insight during the course of the project. The work was funded by Innosuisse as an innovation project under the project number \texttt{54383.1 IP-ICT}.
\end{acks}



\bibliographystyle{ACM-Reference-Format}
\bibliography{main}

\end{document}